\def \abc#1#2#3#4 {\reference#1, {\sl#2}, {\bf#3}, #4}
\def \blank {\lower 5pt\hbox to 0.75in{\hrulefill}}

\def \s{~\rm{s}}
\def \km{~\rm{km}}

\def \K{~\rm{K}}

\documentclass[12pt,preprint]{aastex}

\begin{document}
\small
\setcounter{page}{1}

\begin{center}
\bf
FORMATION OF DOUBLE RINGS AROUND EVOLVED STARS
\end{center}

\begin{center}
Noam Soker$^1$ \\
University of Virginia, Department of Astronomy, P.O.~Box 3818 \\
Charlottesville, VA 22903-0818, USA\\
soker@physics.technion.ac.il \\
\end{center}
$^1$ On sabbatical from the University of Haifa at Oranim, Department of Physics\\
Oranim, Tivon 36006, ISRAEL 


\begin{center}
\bf ABSTRACT
\end{center}

I propose a  scenario for the formation of double-ring systems,
as observed in some planetary nebulae and in the outer rings of SN 1987A.
In this scenario, two jets, one jet on each side of the equatorial plane,
expand into a thin, dense shell.
Such a flow is expected in binary systems where the mass-losing primary
undergoes an impulsive mass loss episode which forms a thin, dense
shell. I assume that a small fraction of that mass is accreted onto 
a companion which blows the jets. 
Each jet accelerates shell's material it hits sideways,
forming a higher density ring. 
Using several simplifying assumptions, I derive an expression for the 
radius of the ring, which depends relatively weakly on the jets and 
impulsive mass loss episode properties. 
This shows that such a scenario is feasible.
If there are several such impulsive mass loss episodes, more double rings
will be formed, as observed in the planetary nebulae MyCn 18 
(the Hourglass nebula). 
Because of the binary interaction and orbital motion, 
the double-ring system is displaced from the symmetry axis of the main
nebula, as observed. 

{\it Subject headings:}
  stars: mass-loss
--- binaries: general 
--- circumstellar mater
--- planetary nebulae: general
--- supernovae: individual (SN 1987A)  

\section{INTRODUCTION}

 Two rings, one on each side of the central star (hereafter termed 
``double rings''), have been reported for several evolved systems. 
 The most famous is the double-ring system (the outer rings) of SN 1987A
(Panagia et al. 1996; Crotts, \& Heathcote 2000; Maran et al.\ 2000,
and references in these papers). 
Double-ring systems have been found in the planetary nebulae (PNe)
Hubble 12 (Welch et al.\ 1999) and He 2-113 (Sahai, Nyman, \& Wootten 2000).
The PN MyCn 18 (the Hourglass nebula; Sahai et al.\ 1999) contains several
pairs of rings, and it is not clear yet if their formation mechanism
is the same as that of double-ring systems. 

 Recently, Hony et al.\ (2001) reported the discovery 
of a double-ring structure around the central star of the $\eta$ Car nebula. 
However, higher resolution observations show that these rings are not real
(Smith et al. 2001b). 
There are other arguments contradicting the claim by
Hony et al. (2001). 
(i) The more prominent ring of $\eta$ Car (to the southwest) 
has a circular appearance, or it is even slightly elongated in the radial
direction from the central star ($20 \mu$m map of Hony et al. 2001). 
However, it is expected that circular rings should look like 
ellipses elongated in the azimuthal direction, as in 
SN 1987A (Crotts \& Heathcote 2000) and the several rings in the 
Hourglass nebula (MyCn 18; Sahai et al.\ 1999). 
(ii) The two rings of $\eta$ Car are projected on the southeast lobe,
which is tilted toward us (Davidson et al.\ 2001).
There are several fainter `blobs' and  `fingers' seen on the other lobe, 
which are generally symmetric with the two rings on the southeast lobe. 
It is quite possible that if the northwest lobe had the same inclination
as the northeast lobe, the blobs and filaments would be brighter.
It is possible, therefore, that the `rings' are a result of some instabilities,
or stochastic mass ejection on the northwest lobe.
(iii) The long-slit spectroscopy used by Davidson et al.\ (2001) to infer 
velocities on the Homunculus covers a portion of the southwest ring.
No peculiar velocities are observed there.  
This adds support to the idea that the `rings' are located on the lobe.
(iv) Another possibility is that the rings are in the orbital plane; 
the bipolar nebula around the luminous blue variable G25.5+0.2 
is similar to the Homunculus of $\eta$ Car, and has a double-nested equatorial
ring system (Clark, Steele, \& Langer 2000).
For these reasons and the evidence from the new observations of 
Smith et al. (2001), I will no longer refer 
to the previously claimed double-ring system in $\eta$ Car.

 Some interesting properties of the double-ring systems are:  
(1) The outer rings of SN 1987 are much more prominent than
any of the other double-ring systems, with their density being much higher
than the density of their surroundings (Burrows et al.\ 1995).
In the other systems, the medium around the rings can be seen,
with a much lower density contrast than that in SN 1987A.
For that reason, and because of the presence of the inner ring in the 
equatorial plane of SN 1987A, it is not clear whether the formation mechanism
of the outer rings of SN 1987A is the same as that of double rings in PNe. 
(2) While in some systems (SN 1987; Hb 12) the line connecting the two
rings' centers (hereafter double-ring axis) is more or less along the major 
(symmetry) axis of the nebula, in He 2-113 this line is highly inclined to 
the longer, presumably the symmetry, axis.
(3) In all systems the double-ring system is displaced from the central star,
i.e., the double-ring axis misses the central star (SN 1987A; He 2-113)
and/or the intensity around the rings is not constant (e.g., He 2-113; Hb 12).
(4) The rings are detected via different emissions bands, e.g., 
the double rings of HB 12 are seen in [Fe II] (Welch et al.\ 1999),
and those of He 2-113 in H$\alpha$. 

Many models were proposed for the formation of the outer rings of
SN 1987A, but most of them were found to have some problems.
 Models based on interacting winds (e.g., Blondin, \& Lundqvist 1993) 
were criticized by Burrows et al.\ (1995) and Meyer (1997), 
while the ionization model of Meyer was criticized by Soker (1999). 
Crotts \& Heathcote (2000), discuss these models, as well as the one based 
on a combination of interacting winds and ionization 
(Chevalier \& Dwarkadas 1995), and the model based on an equatorial pulsed 
mass loss (Soker 1999), and find that no model is capable of explaining all 
properties of the outer rings of SN 1987A.
 A recent model proposed by Tanaka \& Washimi (2002), which is based on 
magnetic tension, has too many assumptions concerning the mass loss 
geometry and magnetic field, and it is not clear whether it 
reproduces the properties of the rings. 

 In the present paper, I explore a different model for the formation 
of a double-ring system, which can also explain several
pairs of rings. 
I examine a short pulse of mass loss from the primary mass-losing star,
which forms a spherical, dense, thin shell.
I assume, then, that a jet hits and interacts with the shell (one jet on
each side of the equatorial plane). 
This type of flow is expected to occur if part of the mass lost to the shell
is accreted by a companion, which then blows the two jets.
The central star of He 2-113 is a Wolf-Rayet type star  (a [WC] star); 
it is possible that an impulsive mass loss episode was connected
with the transition to a [WC] star. 
Jets, one on each side of the equatorial plane (if they are not 
well-collimated, they are called CFW for ``collimated fast wind''), 
which last for a long  time are thought to lead to the formation of 
bipolar lobes (Morris 1987; Soker \& Rappaport 2000; Soker 2002).
Accreting close binary companion can form bipolar PNe 
(Soker \& Rappaport 2000), i.e., those with two polar lobes and an 
equatorial waist between them, while wider companions may form elliptical 
PNe (Soker 2001), i.e., PNe with a large-scale elliptical shape, 
but which can possess small lobes or jet-like features.
There is no sharp boundary between the two groups; the approximate
orbital separation between the binary systems that form bipolar, and 
those which form elliptical, PNe depends on the mass and nature of the 
accretor, and on the properties of the wind (particularly its velocity),
blown by the mass losing star (Soker 2001). 
 
The model is discussed and shown to be feasible in $\S 2$. 
Disks can precess, explaining the inclination of the double-ring 
axis with respect to other structures in the nebula. 
This, and other possible explanations are discussed in $\S 3$.
 In $\S 3$, I also discuss the displacement of the double-ring axis
from the central star.
I summarize the main results in $\S 4$. 
 
\section{THE PROPOSED MODEL}

 I treat a jet expanding into a thin shell, and accelerating the
 material in the shell, both forward and to the sides.
 I assume that the jet is formed by an accretion disk
(Livio 2000), which is formed via mass transfer from an evolved star
into a compact companion.
The scenario can work in principle in cases where the jet is formed 
via a single star mechanism, e.g., Garc\'ia-Segura, \& L\'opez (2000).
 The binary system destroys the axisymmetrical nature of the flow. 
However, to facilitate an analytical treatment, I assume axisymmetry.
I return to this point in $\S 3.2$. 
It is assumed that prior to and after the impulsive mass loss
episode, the mass loss rate from the primary was too low for the formation
of jets (Soker \& Rappaport 2000). 
 The impulsive mass loss rate into the shell is $\dot M_p$,
and that into one jet is $\dot M_j$;
their expansion velocities are $v_p$ and $v_j$, respectively. 
The opening angle of the jet, from its symmetry axis to its edge, is 
$\alpha$, i.e., the jet's head covers a solid angle of 
$2 \pi (1 - \cos \alpha) \simeq \pi \alpha^2$, where I assume $\alpha \ll 1$.
The densities in the shell and jet are 
\begin{equation}
\rho_p = {\frac {\dot M_p}{4 \pi r^2 v_p}}, \qquad {\rm and} \qquad
\rho_j \simeq {\frac {\dot M_j}{\pi r^2 \alpha^2 v_j}},
\end{equation}
respectively, and $r$ is the distance from the center.
The velocity of the jet head, $v_h$, is determined by equating the
slow wind pressure on the jet head with that of the jet material.
 Since thermal pressure can be neglected in the pre-shocked media,
the expression is $\rho_j (v_j-v_h)^2 = \rho_p (v_h-v_p)^2$.
For the mechanism proposed here to work, the jet needs to deposit its
energy in the shell.
I therefore assume that the jet substantially slows down in the shell,
such that $v_h \ll v_j$; hence $v_h$ is approximately given by 
\begin{eqnarray} 
v_h \simeq v_p \left[ 1 +\frac{v_j }{v_p} 
\left( \frac{\rho_j }{\rho_p} \right)^{1/2} \right] \simeq
v_j \left( \frac{\rho_j }{\rho_p} \right)^{1/2}. 
\end{eqnarray}
 The second equality comes from the requirement that the jet be energetic 
enough to form the rings.
This can be seen by substituting typical values,
\begin{eqnarray} 
\frac{v_h}{v_p} \simeq 7.3 
\left( \frac{v_j}{40 v_p} \right)^{1/2}
\left( \frac{\dot M_j}{0.01\dot M_p} \right)^{1/2}
 \left( \frac{\alpha}{10^\circ} \right)^{-1}.
\end{eqnarray}
 This scaling may result, for example, from a companion that accretes 
$\sim 10 \%$ of the mass lost in the pulse and blows $10 \%$ of that 
into each jet, which has an opening angle of $\sim 10 ^\circ$.
In progenitors of PNe 
the shell velocity can be $\sim 10 \km \s^{-1}$, while the jet velocity 
from a main sequence companion can be 
$\sim 200-500 \km \s^{-1}$, and an order of magnitude faster from a white
dwarf companion. 

Because the interaction takes place close to the central star where
densities are high so cooling is fast (Soker 2002), and 
the shell is thin, no long lived hot bubble is formed, 
and the acceleration of the shell material by the pressure 
formed by the interaction lasts for a short time, $t_{\rm acc}$. 
I assume that as the jet interacts with the shell, a region with pressure
equal to the ram pressure of the jet is formed. 
 With the assumption $v_h \ll v_j$ this ram pressure is 
$P \simeq \rho_j v_j^2$. 
The thickness of the shell formed by the impulsive mass loss
event is $d_p=v_p t_p$, where $t_p$ is the duration of the pulse. 
I approximate the segment of the shell in the jet's vicinity
as a plane, with the $z$ coordinate along the symmetry axis, i.e.,
perpendicular to the plane of the shell, and with the $y$ coordinate
as the distance from the symmetry axis in this plane. 
 The mass of the shell per unit length along the $z$ direction and in a 
circular cross section between the symmetry axis and a distance $y$ from it,
is
\begin{equation}
m (y)= \frac {d M}{dz}= \pi y^2 \rho_p. 
\end{equation}
  Let the shell material be accelerated to a transverse velocity
$v_0$ as it leaves the interaction region
at a distance $y_0=r \sin \alpha$ from the symmetry axis.
I assume that the pressure accelerates the shell material from the
symmetry axis ($y=0$) to the jet head's radius $y_0$, with 
a force per unit length of $d F/dz= 2 \pi yP$. 
Conservation of momentum gives 
\begin{equation}
v_0 \simeq \left[ \frac { \pi y_0^2 P }{m(y_0)} \right]^{1/2} 
\delta \simeq \left( \frac{\rho_j}{\rho_p} \right)^{1/2}
 v_j  \delta \simeq  v_h \delta . 
\end{equation} 
The transverse acceleration does not occur along the entire jet's cross
section, hence it is less efficient than assumed. 
This is taken into account in the factor $\delta < 1$. 

The formation of a dense, thin envelope around jets propagating in a 
continuous medium is well established both numerically and analytically
(e.g., Blondin, Fryxell, \& K\"onigl 1990; de Gouveia dal Pino \& Benz 1993;
Cerqueira, de Gouveia dal Pino, \& Herant 1997;
Lee et al.\ 2001; Ostriker et al.\ 2001). 
In the present case, the medium is not continuous, hence
a ring, instead of a thin envelope, is formed around the jet.
Ostriker et al.\  (2001) derive an analytical expression for the shape
of the thin envelope (which they call a ``shell''; not to be confused with
the shell used here).
 Using their equation (22) close to the jet, for a short time after
a parcel of gas leaves the interaction region,
I derive the expression $v_0=(\beta c_s /v_h) v_h$,
where $v_0$ and $v_h$ are the notation used in the present paper,
$\beta \sim 4$ is a factor used by Ostriker et al.\ (2001) in
their expression for the momentum flux, and $c_s$ is their sound speed in the
interaction region.
Their ratio $\beta c_s /v_h$ plays the role $\delta$ plays here.
They find the range of this ratio is $0.2-0.6$, therefore
I take $\delta \simeq 0.2-0.6$, as well. 
Using equations (5) and (3), then, gives 
\begin{eqnarray} 
v_0 \simeq 30 
\left( \frac{v_j}{400 \km \s^{-1}} \right)^{1/2}
\left( \frac{v_p}{10 \km \s^{-1}} \right)^{1/2}
\left( \frac{\dot M_j}{0.01\dot M_p} \right)^{1/2}
\left( \frac{\alpha}{10^\circ} \right)^{-1}
\left( \frac{\delta}{0.4} \right) \km \s ^{-1} . 
\end{eqnarray} 

 The condition for equations (5) and (6) to hold is that the acceleration 
time assumed here, $t_{\rm acc} \sim y_0/v_0$, is shorter than the time it
takes the jet to cross the shell, $d_p/v_h$, which gives the condition
$y_0 \lesssim d_p$. 
 In the other limit, $y_0 \gtrsim d_p$, the acceleration time is
$t_{\rm acc}\sim d_p/v_h$, from which the velocity is derived
\begin{equation}
v_0 \sim v_h \frac{d_p}{y_0} \delta \qquad {\rm for} \qquad  y_0 \gtrsim d_p  .
\end{equation}
However, in this case we also expect $\delta$ to have a lower value.
Hence, the case of a too thin shell and a too wide jet is
not efficient for the proposed mechanism. 
 The situation is more complicated.
 At large distances from the star (large $r$) the jet head cross section is 
larger, reducing the efficiency. 
On the other hand, at larger distances the radiative cooling is longer, 
making the acceleration more efficient (Soker 2002).

  For the compression of rings to higher densities than their environment, 
the flow should start as supersonic $v_0>c$, where $c$ is the sound 
speed in the shell.
Before ionization starts, the material blown by the primary
evolved star is cool, $T<1000 \K$, and mainly atomic, or even molecular,
with a sound speed of $c< 4 \km \s^{-1}$.
 The interaction between the transversely expanding ring and the mass
it sweeps from the cold shell is radiative, and the velocity is determined
by momentum conservation, i.e., $m_0 v_0 \simeq m(y) v(y)$, where
$m_0$ is the mass per unit length pushed by the jet to the sides. 
Not all of the shell mass that is hit by the jet head is pushed to the 
side; some is pushed forward, giving $m_0 < m(y_0)$.
This is parametrized by $m_0=\eta m(y_0)$.
For example, if the mass expelled to the sides comes from the
region $0.8 y_0 < y \leq y_0$, then $\eta \simeq 0.4$.
For $y \gg y_0$ the mass per unit length as given by equation (4) can be used.  
 This gives $v(y)=\eta v_0(y/y_0)^{-2}$.  
Using $y_0 = r \sin \alpha \simeq r \alpha$ gives the distance from the 
symmetry axis where compression stops as 
\begin{eqnarray}
\frac{y}{r} \simeq \left( \frac{v_0}{c} \right)^{1/2} \alpha \eta^{1/2}.
\end{eqnarray} 
With the aid of equation (6) this can be scaled as 
\begin{eqnarray}
\frac{y}{r} \simeq 0.3
\left( \frac{v_j}{400 \km \s^{-1}} \right)^{1/4}
\left( \frac{v_p}{10 \km \s^{-1}} \right)^{1/4}
\left( \frac{c}{4 \km \s^{-1}} \right)^{-1/2}
\left( \frac{\dot M_j}{0.01\dot M_p} \right)^{1/4}
\left( \frac{\alpha}{10^\circ} \right)^{1/2}
\left( \frac{\delta}{0.4} \right)^{1/2} 
\left( \frac{\eta}{0.4} \right)^{1/2}.
\end{eqnarray}

 The following should be noted in regard to the last equation.
\newline
(1) This expression was derived under the assumptions that the
    shell segment from which the ring is formed is a plane, 
    and the radial expansion (from the central star) of the material
     in the shell was neglected.
     Hence, this already approximate expression becomes very crude for
     large values of $y/r$. 
\newline
(2) The derived size of the ring for reasonable parameters in PNe 
  is on the order of the observed size of the rings, with a relatively
  weak dependence on the different parameters.
  The same scaling may hold for the progenitor of SN 1987A during the
  formation of the outer rings, because the star was a red super giant,
  and the companion required by other arguments is a main sequence star
  in the mass range of $0.5 \lesssim M_2 \lesssim 3 M_\odot$ 
(Chevalier \& Soker 1989;  Podsiadlowski, Joss, \& Rappaport 1990; Collins 
et al.\ 1999).
 The outer rings were formed just before the companion entered the common 
 envelope phase with the progenitor of SN 1987A. 
\newline
(3) The ring becomes larger for a larger opening angle $\alpha$
of the jet. 
 However, $\alpha$ can't be too large because the conditions 
$v_o>c$ and $r \alpha \la d_p$ ($d_p$ is the shell width)
should be met. 
Therefore, the present scenario can't account for very wide rings close to
the equatorial plane.
Two such rings are observed in the massive eclipsing binary system RY Scuti
(Smith, Gehrz, \& Goss 2001a; Smith et al.\ 1999), and probably were formed
by the equatorial flow (Smith et al.\ 1999).  
\newline
(4) The total interaction time is equal to the time it takes the jet
to cross the shell $d_p/v_h=t_p(v_p/v_h) \ll t_p$ ($t_p$ is the duration of
the impulsive mass loss event that forms the shell). 
It turns out that the jet can be blown in a very short pulse.
\newline
(5) As the jet breaks out of the outer side of the shell it pushes 
the dense shell's material to the sides and exterior
to the shell (like a pencil punching a hole in paper).
 This may be the explanation for fainter gas extending radially from
the rings in He 2-113 (see fig. 1b of Sahai et al.\ 2000, particularly the
eastern ring). 
We note that ionization fronts also tend to form denser tails
behind dense material (see below). 
\newline
(6) Later evolution is not considered here.
In particular, the central stars in all systems considered here become
hotter, ionize the nebulae around them, and blow fast winds.
Ionization proceeds faster in the low density surroundings of the
ring, heating it to higher temperatures, hence increasing the surrounding 
pressure.
 The higher pressure may further compresses the ring and material in its 
shadow (e.g., Canto et al.\ 1998; Pavlakis et al.\ 2001), increasing 
the density contrast. 
Later, the ionization heats the gas in the ring, which starts to expand,
thus reducing the density contrast between the ring and its environment.  
The fast wind accelerates the lower density environment to high velocities,
while the dense ring lags behind. This increases the density contrast.
     
\section{CHANGING THE SYMMETRY AXIS}
\subsection{Changing the Direction of the Jets}
I consider two mechanisms for the inclination of the line joining the 
centers of the two rings to the symmetry axis of the main nebula, e.g.,
He 2-113 (Sahai et al.\ 2000).
 In the first, the accretion disk is precessing. 
The short life of the jets implies that they can be blown while the 
accretion disk has not get precessed very much. 
In the second mechanism, the impulsive mass loss of the shell was very clumpy.
The rings in He 2-113 are also very clumpy (Sahai et al.\ 2000, fig 1b).
 Because of the clumpy nature of the accreted slow wind, more angular 
momentum may be accreted from one side of the equatorial plane, and
the accretion disk (if formed) will be inclined to the equatorial plane. 
 As was found in the previous section, the mass loss rate into each of
the two jets can be $\sim 0.01$ of that in the impulsive mass loss event
from the primary. 
Because  a fraction of $\sim 0.1-0.2$ from the accreted mass is blown 
into the two jets, only a fraction of $\lesssim 0.1$ 
from the mass lost by the primary needs to be accreted. 
 This part can be clumpy. 
 
  For accretion from a wind, the net specific angular momentum of the
material entering the Bondi-Hoyle accretion cylinder with radius 
$R_a = 2 G M_2/v_r$, i.e., the material having impact parameter $b<R_a$, is 
$j_{BH} = 0.5 (2 \pi / P_o) R_a^2$ (Wang 1981), 
where $P_o$ is the orbital period, and $v_r$ is the relative velocity 
between the wind and the companion of mass $M_2$.
  Livio et al.\ (1986; see also Ruffert 1999) find that the actual
accreted specific angular momentum for high Mach number flows is
$j_a = \eta j_{BH}$, where $\eta \sim 0.1$ and $\eta \sim 0.3$
for isothermal and adiabatic flows, respectively.
 Taking a circular orbit for simplicity, the relative orbital velocity
of the two stars is $v_{\rm orb} = 2 \pi a/P_o$, and taking $\eta =0.2$
gives $j_a \simeq  0.1v_{\rm orb} R_a^2/a$.
 The angular momentum is perpendicular to the equatorial plane, and for 
accretion from a homogeneous wind, the accretion disk around the companion
will be in the equatorial plane.
 To significantly tilt the plane of the accretion disk, I consider
a clumpy (inhomogeneous) wind, where an extra mass $\Delta M$ is accreted
from one side of the plane. 
 Let the wind speed be $v_p$ as before, and assume that near the companion
the extra mass has an impact parameter of $\sim 0.5 R_a$ above the plane.
 The specific angular momentum added by this extra mass, $j_{\rm clump}$, is
\begin{eqnarray} 
\frac{j_{\rm clump}}{j_a} = 5 \left( \frac{a}{R_a} \right)
\left( \frac{v_p}{v_{\rm orb}} \right)
\left( \frac{\Delta M}{M_{\rm acc}} \right),
\end{eqnarray}
where $M_{\rm acc}$ is the total mass accreted to the disk.
It is assumed that the companion accretes only $0.1$ of the mass lost
by the primary. This implies that $a \gtrsim 0.5 R_a$.
Also, for relevant PNe I find $0.3 \lesssim v_p/v_{\rm orb} \lesssim 3$.
By using these values in equation (10) 
I conclude that the clumpy impulsive mass loss, with $\sim 10 \%$ 
more mass blown on one side of the orbital plane in the companion 
direction, can significantly tilt the accretion disk and the 
jets blown by it. 
 This requires that the impulsive mass loss episode, or at least the 
duration of blowing the jets, be much shorter than the orbital period.

\subsection{Displacement from Axisymmetry}

 Several processes can lead to a departure from axisymmetry of nebulae 
formed by binary interaction  (see $\S 1$ of Soker \& Rappaport 2001).  
Here I briefly mention several effects, of which all,
or a subset, of them can be significant in any particular case.  
 The displacement of the rings can manifest itself in the
rings' symmetry axis not crossing the central binary system, 
and/or one side of each ring (but the same side for both rings)
being brighter or larger. 
 
 Each of the two stars have an orbital velocity around the center of mass.
If the impulsive mass loss episode is shorter, or not much longer, 
than the orbital period, then the mass leaves the system with a boost in 
the direction toward which the star moves at that period. 
 This effect can influence the shell or the jet. 
The magnitude of the effect is approximately the ratio of the orbital
to the expansion velocities.  
The shell is likely to acquire a larger departure from axisymmetry
since its expansion velocity is much lower than that of the jets. 

Another effect is due to the companion being displaced from the center of
the shell; the initial shell center is at the location of the primary mass 
losing star (at the moment it blows the shell), while the jets are blown 
somewhat later, and at the location of the binary companion.
 If the interaction between the jets and the shell occurs at a distance $r$,
and the orbital separation is $a$, the magnitude of this effect is 
$\sim a/r$. 
Yet, in another process, the jets blown by the companion can be bent by the 
primary's wind (Soker \& Rappaport 2000). 

I conclude that displacement from axisymmetry is a natural outcome of the
scenario proposed in the present paper, and should occur in most
cases. 

\section{SUMMARY}

 A scenario for the formation of double-ring systems was analytically
examined under several simplifying assumptions. 
In this scenario two jets, one jet on each side of the equatorial plane,
expand into a thin, dense shell.
Such a flow is expected in binary systems where the mass losing primary
undergoes an impulsive mass loss episode which forms a thin, dense,
and more or less spherical, shell, and  
a fraction $\sim 0.01-0.1$ of that mass is accreted onto a companion, 
such that an accretion disk is formed, and the companion blows the jets.
To facilitate the derivation of a simple analytical relation that can 
be easily used, I approximated the segment of the spherical shell 
in and near the interaction region as a plane.
 I neglected the radial expansion of the shell, and assumed that a
pressure equal to the ram pressure of the jet accelerates the shell 
material that is impacted by the jet.
Material is pushed to the sides, emerging from the interaction region
with transverse supersonic speed, $v_0>c$. 
Some shell material is pushed forward (in the radial direction) by the jet.
This reduces the efficiency of the acceleration, and lowers the 
mass of the shell that is hit by the jet and which flows to the sides. 
The reduced acceleration efficiency is represented by a parameter $\delta<1$,
and the lower mass by a factor $\eta <1$. 

The derived expression for the radius of the ring, $y$, is given in 
equation (9), which is scaled with typical winds' properties expected 
in progenitors of PNe.
Equation (9) shows that the size of the ring depends relatively weakly
on the winds' properties and the different physical variables and factors.
Hence, I argue, the simple analytical treatment shows that such
a scenario is feasible.
 The same scenario can account for the double-ring system in $\eta$ Car 
(Hony et al.\ 2001), if real (however, see $\S 1$).
 The double rings in SN 1987A (the outer rings) are very prominent.
 The process studied here by itself can't account for their formation.
 However, ionizing radiation emitted by, and a fast wind blown by, 
the progenitor of SN 1987A as it turned to a blue supergiant, could
account for the high density contrast between the rings and their
surroundings. 
 Three dimensional gas-dynamical numerical simulations are required to
verify the proposed scenario. 
 
This scenario can account for several other properties of 
double-ring systems.
If there are several such impulsive mass loss episodes, more double rings
will be formed. Hence, this scenario may account for the rings
in the PN MyCn 18 (the Hourglass; Sahai et al.\ 1999). 
The inclination of the symmetry axis of the double rings to the symmetry
axis of the main nebula, which is observed in a few systems,
can be explained by either a precessing accretion disk,
or a clumpy (inhomogeneous) wind; accretion from a clumpy wind
may form an accretion disk inclined to the equatorial plane ($\S 3.1$). 
The displacement of the double-ring systems from axisymmetry is observed
in most cases.
The displacement manifests itself in one side of the double-ring system
being brighter or larger, or by their symmetry axis being displaced from
the central star.
  This is expected in the presently proposed scenario.
 Because the impulsive mass loss episode is shorter, or not much longer,
than the orbital period, the shell's center and jets' origin 
are in different locations. 
 Another effect is the orbital motion of the two stars, which results
in a net velocity of the shell and jets centers of mass relative
to the center of mass of the  binary system. 
 The wind from the primary can bend the jets blown by the secondary,
in a process which also causes departure from axisymmetry. 
 
\bigskip

\acknowledgements
I thank Robert Link for helpful discussions, and
Elizabeth Blanton for comments on the original manuscript. 
I was supported by a Celerity Foundation Distinguished Visiting 
Scholar grant at the University of Virginia. 
This research was supported in part by grants from the
US-Israel Binational Science Foundation.




\end{document}